\documentstyle[11pt,amsfonts,fullpage]{amsart}

\newcommand{\bbgin}{\begin}

\def\endproof{\qed\smallskip}
\def\blacksquare{\hbox to .60em{\vrule width .60em height .60em}}

\begin{document}

\title[]{On Stationary Vacuum Solutions to the Einstein Equations}

\author[]{Michael T. Anderson}

\thanks{Partially supported by NSF Grant DMS 9802722}

\maketitle

\setcounter{section}{-1}






\section{Introduction.}
\setcounter{equation}{0}

 A stationary space-time $(M, g)$ is a 4-manifold $M$ with a smooth 
Lorentzian metric $g$, of signature $(-, +, +, +),$ which has a smooth 
1-parameter group $G \approx  {\Bbb R} $ of isometries whose orbits are 
time-like curves in $M$. We assume throughout the paper that $M$ is a 
chronological space-time, i.e. $M$ admits no closed time-like curves, 
c.f. \S 1.1 for further discussion. 

 Let $S$ be the orbit space of the action $G$. Then $S$ is a smooth 
3-manifold and the projection
$$\pi : M \rightarrow  S $$
is a principle ${\Bbb R}$-bundle, with fiber $G$. The chronology 
condition implies that $S$ is Hausdorff and paracompact, c.f. [Ha] for 
example. The infinitesimal generator of $G \approx  {\Bbb R} $ is a 
time-like Killing vector field $X$ on $M$, so that
$${\cal L}_{X}g = 0. $$
The metric $g = g_{M}$ restricted to the horizontal subspaces of $TM$, 
i.e. the orthogonal complement of $<X> \ \subset TM$ then induces a 
Riemannian metric $g_{S}$ on $S$. Since $X$ is non-vanishing on $M$, 
$X$ may be viewed as a time-like coordinate vector field, i.e. $X = 
\partial /\partial t,$ where $t$ is a global time function on $M$. The 
time function $t$ gives a global trivialization of the bundle $\pi$ and 
so induces a diffeomorphism from $M$ to ${\Bbb R} \times S$. The metric 
$g_{M}$ on $M$ may be then written globally in the form
\begin{equation} \label{e0.1}
g_{M} = - u^{2}(dt+\theta )^{2} + \pi^{*}g_{S}, 
\end{equation}
where $\theta $ is a connection 1-form for the ${\Bbb R}$-bundle $\pi$  
and
\begin{equation} \label{e0.2}
u^{2} = -  \langle  X, X \rangle  >  0. 
\end{equation}
The 1-form $\xi $ dual to $X$ is thus given by $\xi  = - 
u^{2}(dt+\theta ).$ The 1-form $\theta$ is uniquely determined by 
$g_{M}$ and the time function $t$, but of course changes by an exact 
1-form if the trivialization of $\pi$ is changed. We point out that 
$(M, g_{M})$ is geodesically complete as a Lorentzian manifold if and 
only if $(S, g_{S})$ is complete as a Riemannian manifold, c.f. Lemma 
1.1.

 The vacuum Einstein field equations on the space-time $(M, g)$ are
\begin{equation} \label{e0.3}
r_{M}  = 0, 
\end{equation}
where $r_{M}$ is the Ricci curvature of $(M, g_{M}).$ 

\medskip

 Stationary vacuum space-times are usually considered as the possible 
final, i.e. time-independent, states of evolution of a physical system, 
in particular isolated physical systems such as isolated stars or black 
holes, outside regions of matter. The most important non-trivial 
example is the Kerr metric, c.f. [W], modeling the time-independent 
gravitational field outside a rotating star.

\medskip

 It is easy to see from the field equations, c.f.(1.4) below, that 
there are no non-flat stationary vacuum solutions of the field 
equations (0.3) whose orbit space is a closed 3-manifold $S$. Hence, we 
will always assume that $S$ is an open 3-manifold.

 Next, it is natural to consider the class of stationary vacuum 
space-times which are geodesically complete. In this respect, 
Lichnerowicz [L, \S 90] proved that any such solution $(M, g)$ for 
which the 3-manifold $(S, g_{S})$ is complete and asymptotically flat 
is necessarily flat Minkowski space.

 The assumption that $S$ is asymptotically flat is very common in 
general relativity in that such space-times serve as natural models for 
isolated physical systems, e.g. stars or black holes. The reasoning 
here is that as one moves further and further away from an isolated 
gravitational source, the corresponding gravitational field should 
decay as it does in Newtonian gravity, giving in the limit of infinite 
distance the empty Minkowski space-time.

 However, mathematically the requirement that $S$ is asymptotically 
flat is a very strong assumption on both the topology and geometry of 
$S$ outside large compact sets. Further, the reasoning above is not at 
all rigorous. It presupposes that a geodesically complete stationary 
solution of the vacuum equations, i.e. a stationary solution without 
sources, is necessarily empty, and so in particular flat. 

 Consider the fact that there are geodesically complete, non-stationary 
vacuum space-times consisting of gravitational waves, c.f. [MTW, \S 
35.9] or [R, \S 8.8] for example. Again, physically, such space-times 
can be considered as idealized limiting configurations at infinite 
distance from radiating sources. Similarly, if there does in fact exist 
a complete non-flat stationary vacuum solution, say $(M_{\infty}, 
g_{\infty}),$ then there could well exist models $(M, g)$ for isolated 
physical systems which are asymptotic to $(M_{\infty}, g_{\infty})$ at 
space-like infinity. For instance, it is not even clear apriori that 
the curvature of a stationary space-time, vacuum outside a compact 
source region, should decay anywhere at infinity.

\medskip

 The first main result of this paper is that in fact there are no such 
non-trivial stationary space-times; this of course places the physical 
reasoning above on stronger footing.
\bbgin{theorem} \label{t 0.1.}
  Let $(M, g)$ be a geodesically complete, chronological, stationary 
vacuum space-time. Then $(M, g)$ is the flat (i.e. empty) Minkowski 
space $({\Bbb R}^{4}, \eta ),$ or a quotient of Minkowski space by a 
discrete group $\Gamma $ of isometries of ${\Bbb R}^{3},$ commuting 
with $G$. In particular, $M$ is diffeomorphic to $S \times {\Bbb R}$, 
$d\theta  =$ 0 and $u =$ const.
\end{theorem}

 This result, together with Lemma 1.1 below implies that if $(M, g)$ is 
a non-flat stationary vacuum space-time, then the orbit space $S$ must 
have a non-empty metric boundary. More precisely, since $(S, g_{S})$ is 
Riemannian, let $\overline S$ denote the metric, (equivalently the 
Cauchy), completion of $S$ and let $\partial S = \overline S \setminus 
S.$ Hence 
\begin{equation} \label{e0.4}
\Sigma  = \partial S \neq  \emptyset  , 
\end{equation}
if $(M, g)$ is not flat.

 In order to avoid trivial ambiguities, we will only consider maximal 
stationary quotients $S$. For example any domain $\Omega $ in ${\Bbb 
R}^{3}$ with the flat metric, $u$ a positive constant, and $\theta  =$ 
0 generates a stationary vacuum solution, (namely a domain in Minkowski 
space). In this case, the metric boundary $\partial\Omega $ is 
artificial, and has no intrinsic relation with the geometry of the 
solution. The solution obviously extends to a larger domain, i.e. all 
of Minkowski space. Thus, we only consider maximal solutions $(S, 
g_{S}, u, \theta ),$ in the sense that the data $(S, g_{S}, u, \theta 
)$ does not extend to a larger domain $(S' , g_{S}' , u' , \theta' ) 
\supset  (S, g_{S}, u, \theta )$ with $u'  > $ 0 on $M'.$ It follows 
that in any neighborhood of a point $q\in\Sigma  = \partial S,$ either 
the metric $g_{S}$ or the connection 1-form $\theta $ degenerates in 
some way, or $u$ approaches 0 in some way, or both.

 Without any further restrictions, the behavior of the data near 
$\partial S$ can be quite complicated; numerous concrete examples of 
this can be found among the axi-symmetric stationary, or even 
axi-symmetric static, i.e. Weyl, solutions; c.f. [A1] for further 
discussion. In particular, singularities, both of curvature type and of 
non-curvature type, may form at the boundary. The horizon $H = \{u = 
0\}$, viewed as a subset of $S$, may or may not be well-defined in this 
generality; of course it corresponds to the locus in $M$ where the 
Killing vector $X$ becomes null. Even when $H$ is well-defined and 
smooth, in general there may be other, possibly singular, parts to 
$\partial S.$

\medskip

 Theorem 0.1 leads to the following apriori estimate on the norm of the 
curvature of a stationary vacuum solution away from the boundary of 
$S$, and on the rate of curvature blow-up on approach to the boundary.
\bbgin{theorem} \label{t 0.2.}
  There is a constant $K <  \infty $ such that if $(M, g)$ is any 
chronological stationary vacuum solution, (not geodesically complete), 
then
\begin{equation} \label{e0.5}
|R_{M}|[x] \leq  K/\rho^{2}[x], 
\end{equation}
where $R_{M}$ is the curvature tensor of $(M, g)$, $[x]$ is the Killing 
orbit through $x\in M$ and $\rho (x) = dist_{g_{S}}([x], \partial S).$ 
The constant $K$ is independent of the data $(M, g)$.
\end{theorem}

 Note that Theorem 0.2 implies Theorem 0.1 by letting $\rho  
\rightarrow  \infty .$ On the other hand, Theorem 0.2 requires Theorem 
0.1 for its proof. In particular, this result shows that if $\partial 
S$ is compact in the completion $\overline S$, then the curvature of 
$(M, g)$ decays at least quadratically w.r.t. the distance from 
$\partial S$.

\medskip

 The contents of the paper are as follows. We discuss some background 
information and preliminary results in \S 1, needed for the work to 
follow. Theorem 0.1 is proved in \S 2 and Theorem 0.2 is proved in \S 3.

 I would like to thank Piotr Chrusciel and Jim Isenberg for useful 
discussions, the referee for pointing out some needed clarifications and
Grisha Perelman for pointing out an error in a previous version of the paper.

\section{Background and Preliminary Results.}
\setcounter{equation}{0}

{\bf \S 1.1.} A stationary space-time $(M, g)$ uniquely determines the 
orbit data $(S, g_{S}, u, \Omega )$ described in \S 0, where $\Omega = 
d\theta$ is the curvature 2-form of the bundle $\pi$ on $S$. 
Conversely, given arbitrary orbit data $(S, g_{S}, u, \Omega )$, $u > $ 
0, satisfying certain equations, (c.f. (1.3)-(1.6) below), there is a 
unique stationary space-time $(M, g)$ in the sense of \S 0, i.e. a 
chronological space-time with a global isometric ${\Bbb R}$-action with 
the given orbit data. 

 Of course, if $(M, g)$ is not chronological, then it will not be 
uniquely determined by the orbit data. One may for instance take a 
${\Bbb Z}$-quotient of $(M, g)$, preserving the orbit data. More 
importantly, if $(M, g)$ is not chronological, then the orbit space $S$ 
may not be a manifold; even if $S$ is a manifold, it may not be 
Hausdorff, c.f. [Ha]. Since the arguments to follow are global on $S$, 
we require that $S$ is globally well-behaved, which is ensured by the 
chronology condition. It is not known for instance if Theorem 0.1 is 
valid without this assumption.

 Recall that a space-time $(M, g)$ is geodesically complete if all 
geodesics in $(M, g)$, parametrized by an affine parameter $s$, are 
defined for all $s\in{\Bbb R} .$ The vertical subspace of $TM$ is the 
subspace spanned by the Killing field $X$ and the horizontal 
distribution ${\cal H}$ is its orthogonal complement in $TM$, defined 
by the metric $g_{M}.$
\bbgin{lemma} \label{l 1.1.}
  A stationary space-time $(M, g_{M})$ is geodesically complete if and 
only if the orbit space $(S, g_{S})$ is geodesically complete.
\end{lemma}
{\bf Proof:}
 Suppose $(M, g_{M})$ is geodesically complete. Let $\gamma $ be a 
geodesic in $S$. Since the projection $\pi : M \rightarrow  S$ is a 
principle fiber bundle, with horizontal spaces ${\cal H} \subset  TM$, 
the geodesic $\gamma $ may be lifted to a horizontal geodesic $\bar 
\gamma$ in $(M, g_{M}),$ with the same parametrization. Since $(M, 
g_{M})$ is complete, $\bar \gamma$ is defined for all values of the 
parameter, and hence so is $\gamma .$

 Conversely, suppose $(S, g_{S})$ is geodesically complete, and hence 
complete as a metric space. Let $\gamma $ be a geodesic in $M$, with 
affine parameter $s$ and tangent vector $T$. Then the projection 
$\sigma  = \pi\circ\gamma $ is a curve in $S$, whose acceleration is 
given by
\begin{equation} \label{e1.1}
\nabla_{V}V = \frac{1}{2}\kappa^{2}\nabla u^{-2} -  \frac{1}{2}\kappa 
L(V). 
\end{equation}
Here $V = d\sigma /ds = \pi_{*}T, \nabla $ is the covariant derivative 
in $(S, g_{S}), \kappa  = < X, T>  =$ const and $L$ is the linear map 
defined by $<L(A), B> X =$ [A, $B]^{v}$ where $A, B$ are horizontal 
vector fields on $M$ and $v$ is the vertical projection, c.f. [T, 
Ch.18.3] for example. Conversely, any curve $\sigma $ satisfying (1.1) 
lifts to a geodesic in $(M, g)$.

 The equations (1.1) form a $2^{\rm nd}$ order system of ODE w.r.t. the 
parameter $s$; note that $L(V)$ is linear in $V$, while $\kappa $ is a 
constant in $s$, depending linearly on $V$. By local existence and 
uniqueness, there exist locally defined solutions $\sigma $ for 
arbitrary initial data $(x, V(x))\in TS.$ Since $S$ is complete, it 
follows that $\sigma $ exists for all values of $s\in{\Bbb R} .$ Hence 
$(M, g)$ is geodesically complete.

{\endproof}

\bbgin{remark} \label{r 1.2.}
  {\rm It is easy to verify that if $(M, g)$ is a stationary, 
(strongly) globally hyperbolic space-time, in the sense that $(M, g)$ 
admits a geodesically complete Cauchy surface $L$, (w.r.t. the induced 
metric), then $(M, g)$ is geodesically complete. The converse issue 
however, i.e. whether a chronological, stationary and geodesically 
complete space-time is necessarily globally hyperbolic, is not clear to 
the author, at least without further assumptions on $u$ and $\theta$. 

  For brevity, we will often say that $(M, g_{M})$ or $(S, g_{S})$ is 
complete instead of geodesically complete.}

\end{remark}

{\bf \S 1.2.} Let $\xi  = - u^{2}(dt+\theta )$ be the 1-form dual to 
the Killing vector $X$, as in \S 0. The twist potential $\omega $ is 
the 1-form on $M$ defined by
\begin{equation} \label{e1.2}
\omega  = \frac{1}{2}*(\xi\wedge d\xi ), 
\end{equation}
It is easily verified that $\omega $ is $G$-invariant, and that it 
descends to a 1-form $\omega $ on the base space $S$. The form $\omega 
$ represents the obstruction to integrability of the horizontal 
distribution in $TM$, and so is related to the curvature 2-form $\Omega 
$ of the connection 1-form $\theta .$ In fact, one easily verifies that
$$2\omega  = - u^{4}*d\theta  = - u^{4}*\Omega ,$$ 
on $(S, g_{S}).$ 

 The vacuum Einstein equations (0.3) on $(M, g)$ are $G$-invariant, and 
so also descend to equations on $S$. The vacuum equations are 
equivalent to the following equations on $(S, g_{S}):$
\begin{equation} \label{e1.3}
r = \frac{1}{u}D^{2}u + 2u^{-4}(\omega\otimes\omega  -  
|\omega|^{2}\cdot  g), 
\end{equation}
\begin{equation} \label{e1.4}
\Delta u = - 2u^{-3}|\omega|^{2}, 
\end{equation}
\begin{equation} \label{e1.5}
div \omega  = 3\langle dlog u, \omega \rangle , 
\end{equation}
\begin{equation} \label{e1.6}
d\omega  = 0. 
\end{equation}
Here $r = r_{S}$ is the the Ricci curvature of $(S, g_{S}), D^{2}u$ is 
Hessian of $u$ on $(S, g_{S})$, $\Delta u = tr_{g_{S}}D^{2}u$ and $log$ 
is the natural logarithm; we refer for instance to [Kr, Ch. 16] for a 
derivation of these equations, (but note that [Kr] does not use the 
factor $\frac{1}{2}$ in (1.2)). The equation (1.3) comes from the pure 
space-like (or horizontal) part of $r_{M},$ the equation (1.4) from the 
vertical part of $r_{M},$ i.e. $r_{M}(X,X),$ while the equations 
(1.5)-(1.6) come from the mixed directions. The equation (1.6) implies 
that $\omega $ is locally exact, i.e. there exists $\phi ,$ the twist 
potential, such that 
\begin{equation} \label{e1.7}
2\omega  = d\phi  
\end{equation}
locally. On the universal cover $\widetilde S$ of $S$, (1.7) holds 
globally. 

 Observe that these equations are invariant under the substitutions 
\begin{equation} \label{e1.8}
u \rightarrow  \lambda u, \omega  \rightarrow  \lambda^{2}\omega , 
\end{equation}
corresponding to $\xi  \rightarrow  \lambda\xi ,$ and $\theta 
\rightarrow \lambda^{-2}\theta$.

\medskip

{\bf \S 1.3.} To prove Theorems 0.1 and 0.2, we will need to study 
sequences of stationary (vacuum) solutions, where all the data $(S, 
g_{S}, u, \omega )$ are allowed to vary. Thus, in effect, we need to 
understand aspects of the moduli space of stationary solutions. For 
this, we will frequently use the following two Lemmas, which will be 
proved together.

\bbgin{lemma} \label{l 1.3}
{\bf (Convergence).}
 Let $(\Omega_{i}, g_{i}, u_{i}, \omega_{i})$ represent data for a 
sequence of solutions to the stationary vacuum equations (0.1). Suppose 
on the domains $(\Omega_{i}, g_{i})$,
\begin{equation} \label{e1.9}
|r_{i}| \leq  \Lambda , \ \ diam \ \Omega_{i} \leq  D, \ \ vol \ 
\Omega_{i} \geq  \nu_{o}, 
\end{equation}
and
\begin{equation} \label{e1.10}
dist(x_{i}, \partial\Omega_{i}) \geq  \delta , 
\end{equation}
for some $x_{i}\in\Omega_{i}$ and positive constants $\nu_{o}, \Lambda 
, D, \delta .$ Then, for any $\varepsilon = \varepsilon(\delta)  > $ 0 
sufficiently small, there are domains $U_{i}\subset\Omega_{i},$ with 
$\varepsilon /2 \leq  dist(\partial U_{i}, \partial\Omega_{i}) \leq  
\varepsilon ,$ and $x_{i}\in U_{i}$ such that a subsequence of the 
Riemannian manifolds $(U_{i}, g_{i}, x_{i})$ converges, in the 
$C^{\infty}$ topology, modulo diffeomorphisms, to a limit manifold $(U, 
g, x)$, with limit base point $x =$ lim $x_{i}.$ 

 Further, the potentials $u_{i}$ and 1-forms $\omega_{i}$ may be 
renormalized by scalars $\lambda_{i},$ as in (1.8), so that they 
converge smoothly to limit potential $u$ and 1-form $\omega .$ The 
limit $(U, g, x, u, \omega)$ represents a smooth solution to the 
stationary vacuum equations.
\end{lemma}

\bbgin{lemma} \label{l 1.4}
{\bf (Collapse).}
 Let $(\Omega_{i}, g_{i}, u_{i}, \omega_{i})$ represent data for a 
sequence of solutions to the stationary vacuum equations (0.1). Suppose 
on the domains $(\Omega_{i}, g_{i})$,
\begin{equation} \label{e1.11}
|r_{i}| \leq  \Lambda , \ \ diam \ \Omega_{i} \leq  D, \ \ vol \ 
\Omega_{i} \rightarrow  0 
\end{equation}
and
\begin{equation} \label{e1.12}
dist(x_{i}, \partial\Omega_{i}) \geq  \delta , 
\end{equation}
for some $x_{i}\in\Omega_{i}$ and constants $\Lambda , D, \delta .$ 
Then, for any $\varepsilon = \varepsilon(\delta)  > $ 0 sufficiently 
small, there are domains $U_{i}\subset\Omega_{i},$ with $\varepsilon /2 
\leq  dist(\partial U_{i}, \partial\Omega_{i}) \leq  \varepsilon $ with 
$x_{i}\in U_{i},$ such that $U_{i}$ is either a Seifert fibered space 
or a torus bundle over an interval. In both cases, the $g_{i}$-diameter 
of any fiber $F$, (necessarily a circle $S^{1}$ or torus $T^{2}),$ goes 
to 0 as $i \rightarrow  \infty ,$ and $\pi_{1}(F)$ injects in 
$\pi_{1}(U_{i}).$

 Consequently, the universal cover $\widetilde U_{i}$ of $U_{i}$ does 
not collapse and hence has a subsequence converging smoothly to a limit 
$(\widetilde U, g, x)$, with $x =$ lim $x_{i}' , x_{i}' $ a lift of 
$x_{i}$ to $\widetilde U_{i}.$ In addition, the limit $(\widetilde U, 
g, x)$ admits a free isometric ${\Bbb R}$-action.

 As above, the potentials $u_{i}$ and 1-forms $\omega_{i},$ after 
possible renormalization by scalars, converge smoothly to limits $u$ 
and $\omega .$ The limit $(\widetilde U, g, x, u, \omega )$ is a smooth 
solution of the stationary vacuum equations, and all data are invariant 
under a free isometric ${\Bbb R}$-action on $\widetilde U.$
\end{lemma}
{\bf Proofs:}
 The proofs of the first parts of Lemmas 1.3 and 1.4 are essentially 
immediate consequences of the well-known Cheeger-Gromov theory on 
convergence and collapse of Riemannian manifolds with bounded 
curvature, c.f. [CG1,2], [Ka], [A3,\S 2] for example. We note that we 
are implicitly using the fact, special to dimension 3, that the full 
curvature is determined by the Ricci curvature. 

 More precisely, under the bounds (1.9)-(1.10), one obtains convergence 
of a subsequence of $\{g_{i}\}$ to a $C^{1,\alpha}$ limit metric $g$ on 
the domain $U$; the convergence is in the $C^{1,\alpha'}$ topology, for 
any $\alpha'  <  \alpha  < $ 1. For a clear introduction to this 
theory, c.f. [P, Ch. 10]. In particular, the bounds (1.9) imply a lower 
bound on the injectivity radius of every point in $U_{i}$; this is 
Cheeger's lemma, c.f. [C], [P, 10.4.5]

 Under the bounds (1.11)-(1.12), the sequence of domains collapses with 
bounded curvature in the sense that the injectivity radius at every 
point in $U_{i}$ tends to 0. This implies that the domains $U_{i}$ 
admit an F-structure, [CG1,2]. In dimension 3, this means that $U_{i}$ 
is topologically a graph manifold, i.e. a union of Seifert fibered 
spaces ($S^{1}$ fibrations over a surface) or torus bundles over an 
interval, glued together along toral boundary components of such, c.f. 
[Ro, \S 3]. A result of Fukaya, c.f. [F, Ch.11,12] and references 
therein, implies that on domains of bounded diameter, i.e. under 
(1.11)-(1.12), for $i$ sufficiently large, the F-structure may be 
chosen to be pure, so that $U_{i}$ itself is either a Seifert fibered 
space or a torus bundle over an interval. The collapse takes place by 
shrinking the fibers, (circles or tori), to points. From the theory of 
Seifert fibered spaces, c.f. [O] or [Ro, Thm. 4.3], the fibers inject 
in $\pi_{1}$ whenever $U_{i}$ is not covered by $S^3$. But this is 
necessarily the case here, since $U_{i}$ is an open domain, (c.f. the 
remark following (0.3). Thus, one may unwrap the collapse by passing to 
covers, for instance the universal cover, that unwind the fibers. This 
ability to unwrap collapse on domains of controlled diameter is special 
to dimension 3.

 It remains to show that the convergence is actually smooth 
$(C^{\infty}),$ and that the limit, in either case of Lemma 1.3 or 1.4, 
is a smooth solution to the stationary vacuum equations. This is done 
by showing that the equations (1.3)-(1.6) form essentially an elliptic 
system and using elliptic regularity. 

 By taking the trace of (1.3) and using (1.4), one derives that
\begin{equation} \label{e1.13}
s = - 6u^{-4}\cdot |\omega|^{2}, 
\end{equation}
where $s$ is the scalar curvature of $(S, g_{S})$, so that (1.4) is 
equivalent to
\begin{equation} \label{e1.14}
\Delta u = \frac{s}{3}u. 
\end{equation}
Since, by hypothesis, the Ricci curvature is uniformly bounded on 
$(\Omega_{i}, g_{i}),$ so is the scalar curvature $s_{i}.$ Now the 
potential functions $u_{i}$ may be unbounded, or converge to 0, in 
neighborhoods of the base points $x_{i}.$ Thus, we renormalize $u_{i}$ 
by setting
\begin{equation} \label{e1.15}
\bar u_{i} = u_{i}/u(x_{i}), 
\end{equation}
so that $\bar u_{i}(x_{i}) =$ 1. The equation (1.14) is of course 
invariant under this renormalization. Moreover, since $u_{i} > $ 0 
everywhere, and since the local geometry of $(\Omega_{i}, g_{i})$ is 
uniformly controlled in $C^{1,\alpha}$ away from $\partial \Omega_{i}$, 
i.e. within $U_{i}$, the Harnack inequality, (c.f. [GT, Thm. 8.20]), 
applied to the elliptic equation (1.14) implies that there is a 
constant $\kappa  > $ 0, independent of $i$, such that
\begin{equation} \label{e1.16}
\kappa  \leq  \frac{sup \bar u_{i}}{inf \bar u_{i}} \leq  \kappa^{-1}; 
\end{equation}
here the $sup$ and $inf$ are taken over $U_{i}$, or more precisely over 
an $\varepsilon /4$ thickening of $U_{i}$. Of course the diameter bound 
in (1.9) or (1.11) is being used here. It then follows from $L^{2}$ 
elliptic theory, c.f. [GT, Thm. 9.11], that the functions $\bar u_{i}$ 
are uniformly bounded in $L^{2,p}(U_{i}), p <  \infty$. Next, as in 
(1.15), we renormalize the twist 1-forms $\omega_{i}$ by
\begin{equation} \label{e1.17}
\bar \omega_{i} = \omega_{i}/(u(x_{i}))^{2}, 
\end{equation}
c.f. (1.8). It then follows from (1.13), (1.15), (1.17) and the uniform 
$L^{\infty}$ bound on $s_{i}$ that the forms $\bar \omega_{i}$ are 
uniformly bounded in $L^{\infty}$ on $U_{i}$.

 Next, to obtain higher regularity, consider the equations (1.5)-(1.6)
$$\Delta\phi_{i} = 3\langle dlog \bar u_{i}, d\phi_{i} \rangle , $$
locally, i.e. in neighborhoods where the twist potential $\phi  = 
\phi_{i}$ is defined; (we omit the overbar from the notation for 
$\phi$). We may add a constant to $\phi_{i}$ and assume 
$\phi_{i}(x_{i}) =$ 0. By the bound on $\bar \omega_{i}$ above, 
$|d\phi_{i}|$ is uniformly bounded, as is $|dlog \bar u_{i}|,$ so by 
elliptic regularity, $\phi_{i}$ is bounded locally in $L^{2,p},$ and 
hence $\bar \omega_{i}$ is uniformly bounded locally in $L^{1,p}$ 
everywhere in $U_{i}$. By (1.13) again, this implies $s_{i}$ is bounded 
in $L^{1,p},$ and so by elliptic regularity applied to (1.14), $\bar 
u_{i}$ is uniformly bounded locally in $L^{3,p}.$ Hence, the right side 
of (1.3) is bounded in $L^{1,p},$ and so the Ricci curvature $r_{i}$ is 
uniformly controlled locally in $L^{1,p}$ everywhere in $U_{i}$. This 
implies that the metrics $g_{i}$ are uniformly controlled in $L^{3,p}$ 
in local harmonic coordinates, c.f. [A3. \S 3] for example. Hence, by 
the Sobolev embedding theorem, the sequence $\{g_{i}\}$ is uniformly 
bounded in $C^{2,\alpha}$, $\alpha < 1$.

 This process may now be iterated inductively to give uniform $C^{k}$ 
control on $\{g_{i}\},$ for any $k <  \infty ,$ away from the boundary, 
as well as uniform $C^{k}$ control on $\{\bar u_{i}\}$ and on $\{\bar 
\omega_{i}\}.$ This proves that the convergence to the limit is in the 
$C^{\infty}$ topology, as well as $C^{\infty}$ convergence to limits 
$\bar u$ and $\bar \omega.$ Since the metrics $g_{i}$ are stationary 
vacuum solutions, it is obvious that the limit $(U, g, \bar u, \bar 
\omega)$ is also.

{\endproof}

 As an application of these results, we prove the following Lemma, 
which shows that a given complete stationary vacuum solution gives rise 
to another one with uniformly bounded curvature.
\bbgin{lemma} \label{l 1.5.}
  Let $(S, g, u, \omega ), g = g_{S},$ represent data for a complete 
non-flat stationary vacuum solution. Then there exists another complete 
non-flat stationary vacuum solution given by data $(S' , g' , u' , 
\omega' )$, $g'  = g_{S'}' ,$ obtained as a geometric limit at infinity 
of $(S, g)$, which has uniformly bounded curvature, i.e.
\begin{equation} \label{e1.18}
|r_{g'}| \leq  1 \ {\rm and} \ |r_{g'}|(y) > 0,
\end{equation}
for some $y \in S'$.
\end{lemma}
{\bf Proof:}
 We may assume that $(S, g)$ itself has unbounded curvature, for 
otherwise there is nothing to prove since (1.18) can then be obtained 
by a fixed rescaling of $(S, g)$ if necessary. Let $\{x_{i}\}$ be a 
sequence in $S$ such that
\begin{equation} \label{e1.19}
|r|(x_{i}) \rightarrow  \infty , \ \ {\rm as} \ \ i \rightarrow  \infty 
. 
\end{equation}
Let $B_{i} = B_{x_{i}}(1)$ and let $d_{i}(x) = dist(x_{i}, \partial 
B_{i}).$ Consider the scale-invariant ratio $(d_{i}^{2}\cdot |r|)(x),$ 
for $x\in B_{i},$ and choose points $y_{i}\in B_{i}$ realizing the 
maximum value of $(d_{i}^{2}\cdot |r|)(x)$ on $B_{i}.$ Since 
$(d_{i}^{2}\cdot |r|)(x)$ is 0 on $\partial B_{i}, y_{i}$ is in the 
interior of $B_{i}.$ By (1.19), we have 
\begin{equation} \label{e1.20}
d_{i}^{2}(y_{i})\cdot |r|(y_{i}) \rightarrow  \infty , \ \ {\rm as} \ \ 
i \rightarrow  \infty  
\end{equation}
and so in particular $|r|(y_{i}) \rightarrow  \infty .$ 

 Now consider the pointed rescaled sequence $(B_{i}, g_{i}, y_{i}),$ 
where 
$$g_{i} = |r|(y_{i})\cdot  g. $$
By construction, $|r_{i}|(y_{i}) =$ 1, where $r_{i}$ is the Ricci 
curvature of $g_{i}.$ This, together with (1.20) and its 
scale-invariance, implies that $\delta_{i}(y_{i})\equiv 
dist_{g_{i}}(y_{i}, \partial B_{i}) \rightarrow  \infty .$ Further, by 
the maximality property of $y_{i},$
\begin{equation} \label{e1.21}
|r_{i}|(x) \leq  |r_{i}|(y_{i})\cdot 
\frac{\delta_{i}(x)}{\delta_{i}(y_{i})} = 
\frac{\delta_{i}(x)}{\delta_{i}(y_{i})} . 
\end{equation}
It follows from (1.20) that $|r_{i}|(x) \leq $ 2, at all points $x$ of 
uniformly bounded $g_{i}$-distance to $y_{i},$ (for $i$ sufficiently 
large, depending on $dist_{g_{i}}(x, y_{i})).$ 

 If the pointed sequence $(B_{i}, g_{i}, y_{i}),$ (or a subsequence), 
is not collapsing at $y_{i},$ i.e. the volume of the unit $g_{i}$-ball 
at $y_{i}$ is bounded below as $i \rightarrow  \infty ,$ then by Lemma 
1.3, $\{(B_{i}, g_{i}, y_{i})\}$ has a subsequence converging, smoothly 
and uniformly on compact subsets, to a limit $(U' , g' , y)$, $y =$ lim 
$y_{i}.$ The limit is a complete stationary vacuum solution, (since 
$\delta_{i}(y_{i}) \rightarrow  \infty ),$ and by the smooth 
convergence, $|r_{g'}| \leq $ 2 everywhere and $|r_{g'}(y)| = 1$, where 
$y$ = lim$y_{i}$. A further bounded rescaling then gives (1.18). The 
limit potential $u$ and twist form $\omega$ are obtained as in Lemma 
1.3.

 On the other hand, suppose this sequence is collapsing at $y_{i},$ so 
that the volume of the unit $g_{i}$-ball at $y_{i}$ converges to 0, (in 
some subsequence). Then by Lemmas 1.3 and 1.4, it is collapsing 
everywhere within $g_{i}$-bounded distance to $y_{i}$, i.e. within 
$(B_{y_{i}}(R), g_{i})$, for any fixed $R < \infty$. For any such $R$, 
if $i$ is sufficiently large, there are domains $U_{i}(R) \subset 
B_{y_{i}}(R)$, with $\partial U_{i}(R)$ near $\partial B_{y_{i}}(R)$ 
w.r.t. $g_{i}$, which are highly collapsed along an injective Seifert 
fibered structure or torus bundle structure on $U_{i}(R)$. Hence the 
universal cover $(\widetilde U_{i}(R), \widetilde g_{i})$ is not 
collapsing. For any sequence $R_{j} \rightarrow \infty$, there is then 
a suitable diagonal subsequence $U_{i_{j}}$ such that the covers 
$\widetilde U_{i_{j}}$ converge smoothly, as above, to a complete 
stationary vacuum solution; again a bounded rescaling then gives (1.18).

{\endproof}

\section{Proof of Theorem 0.1.}
\setcounter{equation}{0}

 Let $(M, g_{M})$ be a complete stationary vacuum solution. As above in 
\S 1.2 and \S 1.3, we will work exclusively on the 3-manifold quotient 
$S$, with data $u$, $\omega $ and $g$ satisfying the field equations 
(1.3)-(1.6). By passing to the universal cover, we may and will assume 
for this section that $S$ is simply connected.

 It is very useful to rewrite the metric $g_{M}$ in (0.1) in the form
\begin{equation} \label{e2.1}
g_{M} = - u^{2}(dt+\theta )^{2} + \frac{1}{u^{2}}\bar g_{S}, 
\end{equation}
where $\bar g_{S}$ is the conformally equivalent metric
\begin{equation} \label{e2.2}
\bar g_{S} = u^{2}\cdot  g_{S} 
\end{equation}
on $S$. Using standard formulas for behavior under conformal changes, 
c.f. [B, Ch. 1J], w.r.t this metric the field equations (1.3)-(1.5) are 
equivalent to:
\begin{equation} \label{e2.3}
\bar r = 2(dlogu)^{2} + 2u^{-4}(\omega )^{2}, 
\end{equation}
\begin{equation} \label{e2.4}
\bar \Delta log u = - 2u^{-4}|\omega|^{2}, 
\end{equation}
\begin{equation} \label{e2.5}
div \omega  = 4\langle dlog u, \omega \rangle , 
\end{equation}
c.f. also [Kr, Ch. 16]. All metric quantities in (2.3)-(2.5) are w.r.t. 
the $\bar g = \bar g_{S}$ metric.

 There are two reasons for preferring $\bar g$ to $g = g_{S}.$ First, 
it is apparent from (2.3) that
\begin{equation} \label{e2.6}
\bar r \geq  0, 
\end{equation}
so that $(S, \bar g)$ has non-negative Ricci curvature. Second, the 
field equations (2.3)-(2.5) are exactly the Euler-Lagrange equations 
for the functional
\begin{equation} \label{e2.7}
S_{eff} = \int_{S}(s -  \frac{1}{2}(\frac{|du^{2}|^{2}+ 
|d\phi|^{2}}{u^{4}}))dV. 
\end{equation}
Here we are using the fact that $S$ is simply connected, so that the 
relation (1.7) holds globally on $S$. This functional is the 
Einstein-Hilbert functional on $G$-invariant metrics on $M$, 
dimensionally reduced to a functional on data $(\bar g, u, \phi )$ on 
$S$, when $g_{M}$ is expressed in the form (2.1). It corresponds to a 
coupling of 3-dimensional gravity to the energy (or $\sigma$-model) of 
the mapping $E = (\phi, u^{2})$ from $S$ to the hyperbolic plane. The 
mapping $E$ is called the Ernst potential and the Euler-Lagrange 
equations (2.3)-(2.5) imply that
\begin{equation} \label{e2.8}
E: (S, \bar g_{S}) \rightarrow  (H^{2}(- 1), g_{-1}) 
\end{equation}
is a harmonic map. Here $H^{2}(- 1)$ is the hyperbolic plane, given as 
the upper half-plane $({\Bbb R}^{2})^{+} = \{(x, y): y >  0\}$, with 
metric
\begin{equation} \label{e2.9}
g_{- 1} = \frac{dx^{2} + dy^{2}}{y^{2}}.
\end{equation}
We refer for instance to [H1] or [H2] for further details and 
discussion on $S_{eff}.$

 From the equation (2.3), we see that
\begin{equation} \label{e2.10}
\bar r = \frac{1}{2}E^{*}(g_{- 1}). 
\end{equation}
In particular, the energy density of $e(E)$ of $E$, given by
$$e(E) = \frac{1}{2}|E_{*}|^{2} $$
satisfies
\begin{equation} \label{e2.11}
\bar s = e(E) = \frac{1}{2}tr_{\bar g} E^{*}(g_{- 1}). 
\end{equation}

 For clarity, we break the proof up at this stage into two steps.

\medskip

{\bf Step I.}
  Assume the metric $(S, \bar g_{S})$ is complete.

  The space $(S, \bar g_{S})$ may or may not have uniformly bounded 
curvature, i.e. possibly after a bounded rescaling,
\begin{equation} \label{e2.12}
|\bar r| \leq  1, 
\end{equation}
everywhere on $S$, where the norm is taken w.r.t. $\bar g_{S}.$ If 
(2.12) holds, then the arguments below are applied to $(S, \bar 
g_{S})$. If instead the curvature of $(S, \bar g_{S})$ is unbounded, 
(and hence $(S, \bar g_{S})$ is not flat), we apply Lemma 1.5 to obtain 
a new {\it non-flat} stationary space-time $(S', \bar g_{S'}, u', 
\omega')$ satisfying (2.12). The arguments below are then applied to 
$(S', \bar g_{S'})$.

  With this understood, we drop the prime from the notation and assume 
that $(S, \bar g_{S})$ satisfies (2.12).

\medskip

 We now apply the well-known Bochner formula, c.f. [EL, (3.12)], to the harmonic Ernst map $E$, to obtain 
\begin{equation} \label{e2.13}
\bar \Delta e(E) = |\bar \nabla DE|^{2} + \langle r_{M}, E^{*}(g_{-1}) \rangle  -  \sum_{i,j=1}^{3}(E^{*}R_{-1})(e_{i},e_{j},e_{j},e_{i}). 
\end{equation}
Here the sign of the curvature tensor for the last term is such that $R_{-1}(X,Y,Y,X)$ is the sectional curvature of $g_{-1}$ for an orthonormal pair $(X,Y)$. We claim that the last two terms in (2.13) are given by
\begin{equation} \label{e2.14}
\langle \bar r, E^{*}(g_{-1}) \rangle  = 2|\bar r|^{2}, 
\end{equation}
\begin{equation} \label{e2.15}
-(E^{*}R_{g_{-1}})(e_{i},e_{j},e_{j},e_{i}) = 4(\bar s^{2} - |\bar r|^{2}) 
\geq 0. 
\end{equation}
The equation (2.14) follows immediately from (2.10). For (2.15), using the 
fact that $g_{-1}$ is of constant sectional curvature $-1$, we have 
$-(E^{*}R_{-1})(e_{i},e_{j},e_{j},e_{i})$ = $g_{-1}(E_{*}e_{i},E_{*}e_{i})
\cdot g_{-1}(E_{*}e_{j},E_{*}e_{j}) - g_{-1}(E_{*}e_{i},E_{*}e_{j})^{2}$. 
Choosing $\{e_{i}\}$ to be an orthonormal basis in $(S, \bar g_{S})$ 
diagonalizing the Ricci curvature $\bar r$, and using (2.10), gives (2.15).

  In particular, the equations (2.13)-(2.15) show that the energy density 
$e(E)$ is a subharmonic function on $(S, \bar g_{S})$.

\medskip

  Since (2.12) holds on $(S, \bar g_{S})$, (2.11) implies that $e(E)$ is 
uniformly bounded above on $(S, \bar g_{S})$. Thus, let $\{x_{i}\}$ be a 
maximizing sequence for $e(F)$, i.e.
\begin{equation} \label{e2.16}
e(F)(x_{i}) \rightarrow  sup \ e(F) < \infty .
\end{equation}
Since the curvature of $(S, \bar g_{S})$ is bounded, and this space is 
complete, it follows 
from elementary properties of the Laplacian that
$$\Delta e(F)(x_{i}) \leq  \varepsilon_{i}, $$
where $\varepsilon_{i} \rightarrow $ 0, as $i \rightarrow  \infty .$ 
However, (2.13)-(2.15) then imply that
$$|\bar r|^{2}(x_{i}) \leq \varepsilon_{i} \rightarrow 0.$$
This of course forces $e(E)(x_{i}) = \bar s(x_{i}) \rightarrow 0$. Since 
$x_{i}$ is a maximizing sequence, this is only possible if
$$e(E) \equiv  0, $$
i.e. $E$ is a constant map. This means that $u =$ const $> $ 0, 
$\phi =$ const, and hence $(M, g)$ is flat. Thus $(M, g)$ is Minkowski space, 
(since $S$ is simply connected). 

  Observe that this argument now implies that the passage to the 
geometric limit $(S', \bar g_{S'})$ at the beginning of Step I was not 
in fact necessary.

\medskip

{\bf Step II.}
 We now remove the assumption that $\bar g$ is complete, by transfering 
the estimates above back to the complete manifold $(S, g_{S})$.

  Exactly as in the beginning of Step I however, since $(S, g_{S})$ is 
complete, if necessary we use Lemma 1.5 first to pass to a non-flat 
geometric limit $(S', g_{S'})$ with uniformly bounded $g'$-curvature, i.e. 
satisfying (1.18). As before, we drop the prime from the notation below.

  Since $\bar g_{S} = u^{2}g_{S},$ we have the following relation between the 
Laplacians of $g_{S}$ and $\bar g_{S}$, c.f. [B, Ch. 1J] for example:
$$\bar \Delta f = u^{-2}\Delta f + u^{-3} \langle du, df \rangle , $$
for any function $f$, where metric quantities on the right are w.r.t. 
$g_{S}$. Setting $f = \bar s$ then gives
\begin{equation} \label{e2.17}
\Delta \bar s = u^{2}\bar \Delta \bar s  -  \langle dlog u, d\bar s 
\rangle . 
\end{equation}
Now the function $\bar s$ may well be an unbounded function on $(S, 
g_{S})$; (in fact the unboundedness may cause the incompleteness of 
$\bar g_{S}).$ 
However, in terms of the metric $g$, we have
\begin{equation} \label {e2.18}
\bar s = u^{-2}(2|dlog u|^{2} + \frac{1}{2}u^{-4}|d\phi|^{2}) \equiv  
u^{-2}\cdot  h,
\end{equation}
where the last inequality defines $h$ and the norms on the right are
w.r.t. $g_{S}$. This follows by taking the trace 
of (2.3).

 Since the curvature of $g_{S}$ is uniformly bounded, i.e. (1.18) holds, 
the same arguments as in the proof of Lemma 1.3-1.4 imply that
\begin{equation} \label{e2.19}
|dlog u|^{2} + u^{-4}|d\phi|^{2} \leq  C , 
\end{equation}
for some $C  <  \infty .$ The estimate (2.19) can also be deduced 
directly from (1.13) and (1.3)-(1.7). Hence, $h$ is uniformly bounded 
above on $(S, g_{S})$. 

 Returning to (2.17), we then have
\begin{equation} \label{e2.20}
\Delta\bar s = \Delta u^{-2}h = u^{-2}\Delta h + h\Delta u^{-2} + 
2\langle du^{-2}, dh \rangle . 
\end{equation}
Now
\begin{equation} \label{e2.21}
\Delta u^{-2} = - 2u^{-3}\Delta u + 6u^{-4}|du|^{2} = u^{-6}|d\phi|^{2} 
+ 6u^{-4}|du|^{2}, 
\end{equation}
where the last equality uses (1.4) and (1.7). Hence, combining 
(2.20)-(2.21), we obtain
$$\Delta h = u^{2}\Delta\bar s -  (u^{-4}|d\phi|^{2} + 
6u^{-2}|du|^{2})h -  2u^{2} \langle du^{-2}, dh \rangle . $$
Substituting (2.17) gives
\begin{equation} \label{e2.22}
\Delta h = u^{4}\bar \Delta \bar s -  (u^{-4}|d\phi|^{2} + 
6u^{-2}|du|^{2})h -  2u^{2}\langle du^{-2}, dh \rangle  -  u^{2}\langle 
dlog u, d\bar s \rangle . 
\end{equation}
Since $\bar s = u^{-2}\cdot  h, d\bar s = - 2u^{-3}hdu + u^{-2}dh,$ and 
so (2.22) becomes
$$\Delta h = u^{4}\bar \Delta \bar s -  (u^{-4}|d\phi|^{2} + 
6u^{-2}|du|^{2})h + 4\langle dlog u, dh \rangle  + 2u^{-2}h|du|^{2} -  
\langle dlog u, dh \rangle , $$
i.e.
$$\Delta h = u^{4}\bar \Delta \bar s  -  (u^{-4}|d\phi|^{2} + 
4u^{-2}|du|^{2})h + 3\langle dlog u, dh \rangle .$$ 
By (2.18) again, the middle term on the right above equals 
$-2h^{2} = -2u^{4}\bar s^{2}$. Hence, we have
\begin{equation} \label{e2.23}
\Delta h - 3\langle dlog u, dh \rangle = u^{4}\bar \Delta \bar s  
-  2u^{4}\bar s^{2}.
\end{equation}
On the other hand, from the Bochner formula (2.13) and (2.14)-(2.15), 
we have
$$\bar \Delta \bar s = |\bar \nabla DE|^{2} + 2|\bar r|^{2} + 4(\bar 
s^{2} - |\bar r|^{2}),$$
where all quantities are w.r.t. the $\bar g$ metric. Substituting this in 
(2.23) then gives
\begin{equation} \label{e2.24}
\Delta h - 3\langle dlog u, dh \rangle = u^{4}|\bar \nabla DE|^{2} + 
2u^{4}(\bar s^{2} - |\bar r|^{2}) \geq 0,
\end{equation}  
where the terms on the left are in the $g$ metric while those on the 
right are in the $\bar g$ metric.

\medskip

 We now basically repeat the argument above in Step I to prove that
\begin{equation} \label{e2.25}
h \equiv 0.
\end{equation}
Thus, recalling from 
(2.19) that $h$ is bounded on $(S, g_{S})$, let $\{x_{i}\}$ be a maximizing 
sequence for $h$. It follows as before that $\Delta h(x_{i}) \leq  
\epsilon_{i}, |dh|(x_{i}) \leq  \epsilon_{i}$ while $|dlog u|(x_{i})$ remains 
uniformly bounded. 

  To prove (2.25), it is most convienient to pass to the limit of the pointed 
sequence $(S, g_{S}, x_{i})$ by use of Lemmas 1.3-1.4. Thus, a subsequence of
$\{(S, g_{S}, x_{i})\}$ converges {\it smoothly}, (passing to covers if 
necessary in the case of collapse), to a complete stationary vacuum solution 
$(S_{\infty}, g_{\infty}, x_{\infty})$. Here the limit potentials $u_{\infty}$
and $\phi_{\infty}$ are limits of the renormalized potentials 
$u_{i} = u/u(x_{i}), \phi_{i} = \phi/u(x_{i})^{2}$. Observe that $h$ and
$dlog u$ are invariant under such renormalizations, as is the right side 
of (2.24) under the changes 
$u \rightarrow u_{i}, \bar g_{S} \rightarrow \bar g_{i} = 
u_{i}^{2}\cdot g_{S}$.

  It follows from these estimates and (2.24), together with the maximum 
principle, that the limit 
$(S_{\infty}, g_{\infty}, x_{\infty}, u_{\infty}, \phi_{\infty})$ satisfies
\begin{equation} \label {e2.26}
h \equiv h_{\infty} = const, \ \ |\bar \nabla DE| = 0, \ \ 
|\bar r|^{2} - \bar s^{2} = 0,
\end{equation}
where $\bar g_{\infty} = u_{\infty}^{2} \cdot g_{\infty}$ and 
\begin{equation} \label{e2.27}
h_{\infty} = sup_{S}h.
\end{equation} 
To see that $h_{\infty} = 0$, (2.26) and (2.10) imply
that $\bar \nabla \bar r = 0$, i.e. the Ricci curvature $\bar r_{\infty}$ 
of $\bar g_{\infty}$ is parallel. By the Bianchi identity this implies that
the scalar curvature $\bar s_{\infty}$ of $\bar g_{\infty}$ is constant.
Since $h = h_{\infty}$ is constant, (2.18) shows that $u_{\infty}$ is also
constant on $(S_{\infty}, g_{\infty})$. Hence by (2.4) on 
$(S_{\infty}, g_{\infty})$, it follows that $d\phi_{\infty} = 0$. By the
definition of $h$ in (2.18), this of course gives $h_{\infty} \equiv 0$,
which by (2.27) gives (2.25).

  The equation (2.25) means that $u$ is a constant function and 
$\omega  =$ 0, so that $d\theta  =$ 0. It follows that 
$(S, g_{S})$ and $(M, g_{M})$ are both flat, which proves the result.

{\endproof}

\section{Proof of Theorem 0.2.}
\setcounter{equation}{0}

 The following result gives Theorem 0.2 essentially as an immediate 
corollary. The proof is a standard consequence of the global result in 
Theorem 0.1, together with the control on moduli of stationary vacuum 
solutions given in Lemmas 1.3 and 1.4.
\bbgin{theorem} \label{t 3.1.}
  Let $(M, g_{M})$ be a stationary vacuum solution, with orbit data 
$(S, g_{S}, u, \theta ),$ and $U \subset\subset  S$ a domain with 
smooth boundary, so that $u > $ 0 on $\overline U.$ Then there is an 
(absolute) constant $K <  \infty ,$ independent of $(M, g_{M})$ and 
$U$, such that for all $x\in U,$
\begin{equation} \label{e3.1}
|r_{S}|(x) \leq  \frac{K}{\rho (x)^{2}}, 
\end{equation}
where $\rho (x) = dist_{g_{S}}(x, \partial U).$
\end{theorem}
{\bf Proof:}
 The proof is by contradiction. Thus, assume that (3.1) does not hold. 
Then there are stationary vacuum solutions $(M_{i}, g_{M_{i}}),$ with 
orbit data $(S_{i}, g_{S_{i}}, u_{i}, \omega_{i}),$ smooth domains 
$U_{i} \subset\subset  S_{i}$ on which $u_{i} > $ 0 and points 
$x_{i}\in U_{i}$ such that
\begin{equation} \label{e3.2}
\rho^{2}(x_{i})|r_{i}|(x_{i}) \rightarrow  \infty , \ \ {\rm as} \ \ i 
\rightarrow  \infty . 
\end{equation}
Let $\rho_{i} = \rho (x_{i}).$ Since it may not be possible to choose 
the points $x_{i}$ so that they maximize $|r_{i}|$ (over large 
domains), we shift the base points $x_{i}$ as follows; compare with the 
proof of Lemma 1.5. Choose $t_{i}\in [0,\rho_{i})$ such that
\begin{equation} \label{e3.3}
t_{i}^{2}  sup_{B_{x_{i}}(\rho_{i}-t_{i})} |r_{i}|   =   sup_{t\in 
[0,\rho_{i})} t^{2} \cdot    sup_{B_{x_{i}}(\rho_{i}- t)}|r_{i}| 
\rightarrow  \infty , \ {\rm as} \ i \rightarrow  \infty , 
\end{equation}
where the last estimate follows from (3.2), (set $t = \rho_{i}).$ Let 
$y_{i}\in B_{x_{i}}(\rho_{i}-t_{i})$ be points such that
\begin{equation} \label{e3.4}
|r_{i}|(y_{i}) = sup_{B_{x_{i}}(\rho_{i}-t_{i})}|r_{i}|. 
\end{equation}
Further, setting $t = t_{i}(1-\frac{1}{k}), k > $ 1, in (3.3), one 
obtains the estimate
\begin{equation} \label{e3.5}
t_{i}^{2}|r_{i}|(y_{i}) \geq  t_{i}^{2}(1-\frac{1}{k})^{2}   \cdot    
sup_{B_{x_{i}}(\rho_{i}- t_{i}(1-\frac{1}{k}))}|r_{i}|  \geq    
t_{i}^{2}(1-\frac{1}{k})^{2}  \cdot     sup_{B_{y_{i}}(t_{i}/k)}|r_{i}| 
,   
\end{equation}
so that
\begin{equation} \label{e3.6}
   sup_{B_{y_{i}}(t_{i}/k)}|r_{i}|  \leq  (1-\frac{1}{k})^{-2} 
|r_{i}|(y_{i}), 
\end{equation}

 Now rescale or blow-up the metric so that $|\widetilde r_{i}|(y_{i}) 
=$ 1 by setting $\widetilde g_{i} = |r_{i}|(y_{i})\cdot  g,$ and 
consider the pointed sequence $(U_{i}, \widetilde g_{i}, y_{i}).$ We 
have 
\begin{equation} \label{e3.7}
|\widetilde r_{i}|(y_{i}) = 1,  
\end{equation}
and by (3.3) and scale invariance,
\begin{equation} \label{e3.8}
dist_{\widetilde g_{i}}(y_{i}, \partial U_{i}) \rightarrow  \infty , \ 
\ {\rm as} \ \ i \rightarrow\infty . 
\end{equation}
Also, (compare with (1.21)), it follows from (3.6) that
\begin{equation} \label{e3.9}
|\widetilde r_{i}|(x) \leq  C(dist_{\widetilde g_{i}}(x, y_{i})). 
\end{equation}
We also normalize $u$ by setting
\begin{equation} \label{e3.10}
\tilde u_{i}(x) = \frac{u(x)}{u(y_{i})}, 
\end{equation}
and note that $\tilde u_{i} > $ 0 on $U_{i}.$

 We may now apply Lemmas 1.3 and 1.4, exactly as in the proof of Lemma 
1.5 to conclude that a subsequence of the pointed sequence $(U_{i}, 
\widetilde g_{i}, \tilde u_{i}, \tilde \omega_{i}, y_{i})$ converges in 
the $C^{\infty}$ topology on compact subsets, to a limit stationary 
vacuum solution $(U_{\infty}, \widetilde g_{\infty}, \tilde u_{\infty}, 
\tilde \omega_{\infty}, y)$, which is complete and satisfies $\tilde 
u_{\infty} > $ 0 everywhere. Here, one must pass to the universal cover 
in case of collapse, as in Lemma 1.4, and the potential $\tilde u_{i}$ 
and 1-form $\tilde \omega_{i}$ are normalized so that $\tilde 
u_{i}(y_{i}) =$ 1 and $|\tilde \omega_{i}(y_{i})|$ is bounded.

 By Theorem 0.1, $\widetilde g_{\infty}$ must be flat, $\tilde 
u_{\infty}$ constant and $d\tilde \omega_{\infty} =$ 0. However, the 
smooth convergence of the sequence $(U_{i}, \widetilde g_{i})$ 
guarantees that the equality (3.7) passes to the limit, contradicting 
the fact that $\widetilde g_{\infty}$ is flat.

{\endproof}

 As in the proof of Lemmas 1.3 and 1.4, it follows from (3.1) that
\begin{equation} \label{e3.11}
|dlog u|(x) \leq  \frac{K}{\rho (x)}, 
\end{equation}
and
\begin{equation} \label{e3.12}
u^{-2}|\omega|(x) \leq  \frac{K}{\rho (x)}. 
\end{equation}
Combining the estimates (3.1) and (3.11)-(3.12), one obtains the same 
bound on the full curvature tensor $R_{M}$ of $(M, g)$. 

 Note that since $K$ is independent of the domain $U$, (3.1) holds for 
$\rho $ the distance to the boundary $\Sigma $ of $S$, even if $\Sigma 
$ is singular. To see this, just apply Theorem 3.1 to a smooth 
exhaustion $U_{j}$ of $S$, with $\partial U_{j}$ converging to 
$\partial S$ in the Hausdorff metric on subsets of $(S, g_{S})$. In 
particular, these results together prove Theorem 0.2. 

{\endproof}

We note that elliptic regularity further implies that, for any $j \geq 
$ 1,
\begin{equation} \label{e3.13}
|\nabla^{j}R_{M}|(x) \leq  \frac{K(j)}{\rho^{2+j}(x)},  |\nabla^{j}log 
u|(x) \leq  \frac{K(j)}{t^{j}(x)}. 
\end{equation}

 Theorem 0.2, when combined with Lemmas 1.3 and 1.4, shows that the 
moduli space of stationary vacuum solutions is apriori well-controlled 
away from the boundary $\Sigma = \partial S$. Thus, away from the 
boundary, sequences of such metrics either have a smoothly convergent 
subsequence, or they collapse, in which case the universal covers have 
a convergent subsequence.

 Theorems 0.1 and 0.2 give new proofs of similar results for static 
vacuum solutions in [An2, Thm. 3.2].  Similarly, in work to follow, we 
plan to generalize the results on the asymptotic structure of static 
vacuum space-times in [A1] to stationary space-times as well as consider
the Riemannian analogues of these questions.

\bibliographystyle{plain}

\bigskip

\begin{center}
October 1999/March 2000
\end{center}
\medskip
\address{Department of Mathematics\\
S.U.N.Y. at Stony Brook\\
Stony Brook, N.Y. 11794-3651}\\
\email{anderson@@math.sunysb.edu}

\end{document}